\documentclass[aps,pre,reprint,superscriptaddress,showpacs]{revtex4-1}
  \usepackage{graphicx}
  \usepackage{color}
  \usepackage{bm}

\newcommand{\newc}{\newcommand}
\newc{\bb}{\begin{eqnarray}}
\newc{\ee}{\end{eqnarray}}
\newc{\nn}{\nonumber}
\newc{\D}{{\rm d}}
\newc{\kt}{\rangle}
\newc{\br}{\langle}
\newc{\ld}{\lambda}
\newc{\e}{{\rm e}}

\begin{document}

\title{Overdamped limit and inverse friction expansion for the Brownian motion in an inhomogeneous
 medium}

\author{Xavier Durang}
\affiliation{School of Physics, Korea Institute for Advanced Study, Seoul 130-722, Korea}
\author{Chulan Kwon}
\affiliation{Department of Physics, Myongji University, Yongin, Gyeonggi-Do 449-728, Korea}
\author{Hyunggyu Park}
\affiliation{School of Physics, Korea Institute for Advanced Study, Seoul 130-722, Korea}

\date{\today}

\begin{abstract}
We revisit the problem of the overdamped (large friction) limit of the Brownian dynamics in an inhomogeneous medium characterized by
a position-dependent friction coefficient and a multiplicative noise (local temperature)
in one space dimension. Starting from the Kramers
equation and analyzing it through the expansion in terms of eigenfunctions of a quantum harmonic oscillator,
we derive analytically the corresponding Fokker-Planck equation in the overdamped limit.
The result is fully consistent with the previous finding by Sancho, San Miguel, and D\"urr \cite{Sanc82}.
Our method allows us to generalize the Brinkman's hierarchy, and thus it would be straightforward to obtain
higher-order corrections in a systematic inverse friction expansion without any assumption.
Our results are confirmed by numerical simulations for simple examples.

\end{abstract}

%02.50.-r	Probability theory, stochastic processes, and statistics
%05.10.Gg	Stochastic analysis methods (Fokker-Planck, Langevin, etc.)
%05.40.-a	Fluctuation phenomena, random processes, noise, and Brownian motion
%66.10.C-	Diffusion and thermal diffusion (for osmosis in biological systems, see 82.39.Wj in physical chemistry; for cellular transport, see 87.16.dp and 87.16.Uv in biological physics)
%05.40.Jc	Brownian motion
\maketitle

The stochastic differential equation (SDE) with a multiplicative noise always brings a basic question about what is the correct choice in representing the noise integration, so called the noise calculus with various types: Ito, Stratonovich, anti-Ito, and others. One consensus is that the noise calculus itself is a part of the problem that should be provided experimentally or theoretically prior to inferring the SDE~\cite{kampen81}.

The Brownian motion of a colloidal particle suspended in a spatially inhomogeneous medium is such an example.
The medium inhomogeneity can be characterized, in general, by space dependence of the friction coefficient  and the local temperature (or the diffusion coefficient). The naive Langevin description in the overdamped (large friction) limit
led to the SDE with a multiplicative noise, which
raised a question of the noise calculus choice, the so-called {\em Ito-Stratonovich dilemma}. However, it is clear that the corresponding underdamped Langevin equation does not depend on the noise calculus, thus the overdamped limit should not depend on it, either.

This dilemma was settled down thirty years ago by Sancho,  San Miguel, and D\"urr (SSMD)~\cite{Sanc82} for most general cases. They successfully integrated out the fast variable (velocity) of the underdamped Langevin equation in the large friction limit by the so-called adiabatic elimination procedure, an extended version of the work done by Haken \cite{Hake78}.
Interestingly, their results do not correspond to any choice of the noise calculus in the naive Langevin description, in general,
except simple cases. However, this derivation is quite involved and mixes up the Langevin equation approach with the Fokker-Planck type description. And their results have never been tested against numerical simulations. These might cause some confusions, which triggered several recent works on this already resolved Ito-Stratonovich dilemma~\cite{Volp10,Ao07,Mann11-12,Sanc11,Yang13,Kuni}.

For a simpler case with a constant friction coefficient (still space-dependent local temperature), the overdamped limit was rigorously derived by
the Fokker-Planck equation approach~\cite{Mats00} and also by the Langevin equation approach at the level of a single realization~\cite{Sekimoto}. This case turns out to correspond to the naive Langevin description with the Ito calculus. The other simpler case with a space-dependent friction coefficient and a constant temperature was also studied
and the Fokker-Planck equation in the overdamped limit was rigorously derived~\cite{titulaer,risken81}, which is equivalent
to the naive Langevin description with the anti-Ito calculus. The overdamped limit for more general cases was rederived
later by a singular perturbation theory~\cite{vKampen} and also by the Chapman-Enskog procedure~\cite{Widd89} with higher-order corrections in the large friction
limit.
%However, the same type of derivation was not possible
%for more general cases, prior to our present work. It should be noted that,  very recently,
%Yang and Ripoll~\cite{Yang13} succeeded to yield the correct overdamped limit for general cases (consistent with the SSMD result),
%through the Langevin equation approach by assuming the macroscopic force balance condition for the averaged quantities.

In this study, we take the standard Fokker-Planck approach to rederive the overdamped limit and
present a systematic inverse friction expansion rigorously for general cases without any assumption.
We start with the Kramers equation for the underdamped Langevin equation, which is independent of the noise calculus. With some operator transformations, we obtain the time-dependent probability
distribution function in terms of the eigenfunctions of the quantum harmonic oscillator, similar to the method employed by
Risken~\cite{Risken89}. Eventually, we generalize the Brinkman's hierarchy~\cite{Brink} which allows for a systematic expansion of the Kramers equation in the large friction limit at any order of the inverse of the friction coefficient. Keeping up to the first order, we find the same overdamped Fokker-Planck equation as in the SSMD. The next order calculation naturally yields the high-order corrections
obtained by Widder and Titulaer (WT)~\cite{Widd89}.
We may emphasize that, thanks to this hierarchy, we obtain a more general time-dependent differential equation for the probability distribution function, which is valid even for very early transient regime.
In order to test the robustness of the first order approximation, we present the numerical simulation results for simple examples, which are in excellent accord with our analytic results. Finally, we show that the overdamped limit is generally equivalent to the mass zero limit in one space dimension, which was previously known only when the Einstein relation (constant temperature) holds~\cite{Ao07}.

We consider the underdamped Langevin equation for one-dimensional Brownian motion of a colloidal particle in an inhomogeneous medium,
which is described by the second order SDE as
\begin{equation}
m\ddot{x}=-\gamma(x) \dot{x}+f(x)+g(x)\xi(t)~,
\label{2ndSDE}
\end{equation}
where $\gamma(x)$ is the friction coefficient and $g(x)$ the noise strength, both of which depend on position $x$. $\xi(t)$ is a white noise satisfying $\langle\xi(t)\xi(t')\rangle=2\delta(t-t')$. Even though the noise is multiplicative (position-dependent),
the choice of the noise calculus is meaningless because the stochastic noise $g(x)\xi (t)$ directly  affects velocity variation rather than position variation in the second order SDE. This will be clear in the probability description, i.e.~Kramers equation later.

Naive approach to the overdamped limit begins with neglecting the inertia term in the left hand side of the equation,
because the overdamped limit is defined in the regime of $\gamma\gg (\Delta t)^{-1}$ with coarse-graining time scale
$\Delta t$. Then, we may end up with
\begin{equation}
\dot{x}=\frac{f(x)}{\gamma (x)}+\frac{g(x)}{\gamma (x)}\xi (t).
\label{convLang}
\end{equation}
However, this equation depends on the noise calculus because the stochastic noise here directly
changes the position instantly, so it will be crucial when the noise strength $g(x)/\gamma (x)$ should be evaluated
during the integration of the equation over time interval $\Delta t$. This dependence of the noise calculus causes the Ito-Stratonovich dilemma. Therefore, the above naive overdamped Langevin equation can not be a correct one to describe the overdamped limit
of the noise-calculus-independent underdamped equation. For its correct description, one should carefully take the proper large $\gamma$ limit, in particular, for the noise-induced drift force and then integrate out the velocity (fast) degree of freedom in the underdamped equation.

For later convenience, we first discuss the noise-calculus dependence of Eq.~(\ref{convLang}). By integrating it during time interval $[t,t+\Delta t]$, one can get the equation for $\Delta x\equiv x(t+\Delta t)-x(t)$ as
\begin{equation}
\Delta x=\left(\frac{f}{\gamma}+2
\alpha\frac{g}{\gamma} \left(\frac{g}{\gamma}\right)'\right)\Delta t +\left(\frac{g}{\gamma}\right)_{\textrm I} \Delta W~,
\label{convSDE}
\end{equation}
where $\Delta W=\int_t^{t+\Delta t}ds~ \xi(s)$ is called the Wiener process, satisfying $\langle \Delta W\rangle=0$, $\langle (\Delta W)^2\rangle=2\Delta t$. The noise calculus parameter $\alpha\in[0,1]$ specifies when the noise amplitude function
$h=g/\gamma$ is
evaluated, such that $\int_t^{t+\Delta t} ds~ h(x(s))\xi(s)$ is replaced by $h(x^*)\Delta W$ with an intermediate value  $x^*=(1-\alpha)x(t)+\alpha x(t+\Delta t)$. Various noise calculi depend on $\alpha$; Ito ($\alpha=0$), Stratonovich ($\alpha=1/2$),
anti-Ito  or isothermal ($\alpha=1$). Employing the Taylor expansion of the noise amplitude function
$h(x^*)$ and the subsequent iteration procedure, the stochastic term can be decomposed into last two terms as above.
The subscript $\textrm I$ (Ito) in the last term indicates that the noise amplitude function should be
evaluated at the initial time $t$ for the Wiener process. The second term is the additional drift force induced by the
noise calculus where the superscript $'$ represents the derivative as $h'=\partial h/\partial x$. Note that this term vanishes when the noise amplitude function $h(x)$ is independent of position $x$.

Following the standard procedure involving the Kramers-Moyal coefficients~\cite{kampen81,Risken89}, it is easy to derive the corresponding Fokker-Plank
equation for the probability distribution function  $P(x,t)$ as
\begin{equation}
\label{convFP}
 \frac{\partial P(x,t)}{\partial t}
=\frac{\partial}{\partial x} \left[ -\frac{f}{\gamma} -\alpha\left(\frac{T}{\gamma}\right)'+
\frac{\partial}{\partial x}\frac{T}{\gamma} \right]P(x,t)
\end{equation}
where the {\em local temperature} $T(x)$ is defined by $T(x)\equiv g^2(x)/\gamma(x)$, called the generalized
Einstein relation.

Now, we return to the underdamped Langevin equation, Eq.~({\ref{2ndSDE}). It is well known that the corresponding probability
evolution (Kramers) equation  can be written as~\cite{Risken89}
\bb \label{Kramer.rev.irr}
\partial_t P(x,v,t) = \left(L_{rev}+L_{irr}\right) P(x,v,t)
\ee
with
\bb\label{rev.irr}
L_{rev} &=& -v \partial_x - (f/m) \partial_v~, \nn \\
L_{irr} &=& (\gamma/m) \partial_v \left[ v+ (T/m) \partial_v\right]~,
\ee
where $\partial_y\equiv \partial/\partial y$ ($y=t,v,x$).
The {\em reversible} operator $L_{rev}$ describes the deterministic motion, while
the {\em irreversible} operator $L_{irr}$ describes the thermal stochastic motion.
As discussed before, there is no dependence on the noise calculus in the Kramers equation,
in contrast to Eq.~(\ref{convFP}).

From now on, we set $m=1$ for simplicity~\cite{exp}.
It is convenient to put $L_{irr}$ into a Hermitian form via a similarity transformation, using
the stationary solution of $L_{irr}$ as $P_{irr}^{s} (v,T) = (2\pi T)^{-1/2}\e^{-v^2/2T}$~\cite{Risken89}.
Then, the Hermitianized operator $\bar{L}_{irr}$ is given as
\bb
\bar{L}_{irr} &=& [{P_{irr}^{s}}]^{-1/2} L_{irr} [{P_{irr}^{s}}]^{1/2}\nn \\
 &=& \gamma\left(T\partial^2_v -\frac{v^2}{4T} +\frac{1}{2}\right)~,
\ee
which is identical to the Hamiltonian operator of a quantum harmonic oscillator.
Introducing the lowering and raising ladder
operators $b$  and $b^\dagger$, we get
\bb
\bar{L}_{irr}= -\gamma b^\dagger b
\ee
with
\bb
b = \sqrt{T}\partial_v  + \frac{v}{2\sqrt{T}}~, \quad
b^\dagger = -\sqrt{T}\partial_v  + \frac{v}{2\sqrt{T}}
\ee
Then, the orthonormal eigenfunctions of $\bar{L}_{irr}$ are given by
\bb
\psi_0(v,T) &=& \left(2\pi T\right)^{-1/4} \exp\left[-{v^2}/{(4T)}\right]=[{P_{irr}^{s}}]^{1/2}~, \nn \\
\psi_n(v,T) &=& b^\dagger \psi_{n-1}(v,T)/\sqrt{n}\nn \\
            &=& \psi_0(v,T)H_n(v/\sqrt{2T})/\sqrt{n!2^n}
\ee
where $H_n$ are the Hermite polynomials ($n=1,2,\ldots$). Note that these eigenfunctions depend
on position $x$ through $T(x)$.
For the operator $L_{rev}$, the same procedure gives
\bb
\bar{L}_{rev} &=& \psi_0(v,T)^{-1} \left(-v \partial_x -f\partial_v\right)\psi_0(v,T)
\\ \nn
&=& -\psi_0^{-1} \partial_x \sqrt{T} \psi_0 (b+b^{\dagger}) +\frac{f}{\sqrt{T}}b^\dagger
\ee

With these transformed operators, Eq.~(\ref{Kramer.rev.irr}) obviously becomes
\bb
\partial_t \bar{P}(x,v,t) = \left(\bar{L}_{rev}+\bar{L}_{irr}\right) \bar{P}(x,v,t)
\ee
with $\bar{P}(x,v,t) = \psi_0^{-1} P(x,v,t)$. It is convenient to decompose the distribution function
in terms of  $\{\psi_n\}$ as
\bb\label{expan}
\bar{P}(x,v,t) = \sum_{n=0}^{\infty} c_n(x,t) \psi_n(v,T)~.
\ee
The transformed operators act on $\bar{P}$ in the following way
\bb
\bar{L}_{irr}\bar{P}&=& -\gamma \sum_{n=0}^{\infty} n \; c_n(x,t) \psi_n(x,T) \\ \nn
\bar{L}_{rev}\bar{P}&=& -\sum_{n=0}^{\infty}\left( [Dc_n] b \psi_{n} +  [\hat{D}c_n] b^\dagger \psi_{n} \right.\\ \nn
                    & & \left.+ c_n \psi_0^{-1} \partial_x \sqrt{T} \psi_0 (b+b^\dagger)\psi_{n} \right)
\ee
where $D = \sqrt{T} \partial_x$, $\hat{D} = \sqrt{T} \partial_x -f/\sqrt{T}$ and $[\cdots]$ means that the operator
acts only inside. When $T$ is a constant, $\{\psi_n\}$ is independent of $x$ and the last term of $\bar{L}_{rev}\bar{P}$
drops out, which simplifies the algebra.

For general $T(x)$, a straightforward algebra
yields with the help of the Hermite polynomial recurrence relation property, $H_n'/H_n=\sqrt{2n} ~\psi_{n-1}/\psi_n$, that
\bb \label{Pbar}
\partial_t \bar{P} &=& -\sum_{n=0}^{\infty} \left(\gamma n c_n + [Dc_n] b  +  [\hat{D}c_{n}] b^\dagger \right. \nn \\
& & \left.+ \left(\sqrt{T}\right)' c_n (b+b^\dagger)b^\dagger(b+b^\dagger) \right)\psi_{n}~.
\ee
From this equation, one can easily extract the hierarchy of the expansion coefficients $c_n(x,t)$ as
\bb \label{Coeff}
\partial_t c_n &=& -\gamma n c_n -(n+1)^{1/2} Dc_{n+1} - n^{1/2} \hat{D}c_{n-1} \nn \\
& & - \left(\sqrt{T}\right)'\left( (n+1)^{3/2} c_{n+1} + 2n^{3/2} c_{n-1} \right. \\
& & \qquad\qquad \left. + \sqrt{n(n-1)(n-2)} c_{n-3} \nn \right)~,
\ee
which is a generalized version of the Brinkman's hierarchy~\cite{Brink,Risken89}.
We emphasize that all results are rigorous without any approximation up to now.

Now, we take the overdamped limit of $\gamma \gg (\Delta t)^{-1}$, in such a way that $\partial_t c_n$ is neglected in comparison with $\gamma c_n$ for $n\geq1$ in Eq.~({\ref{Coeff}). Considering the remaining terms in the order of
the power of $\gamma^{-1}$, one can easily show that $c_n\sim O(\gamma^{-n})$ for $n=0,1,2$ and
$c_n\sim O(\gamma^{-(n-2)})$ for $n\geq 3$. Note that the last term proportional to $c_{n-3}$ in Eq.~({\ref{Coeff}) makes $c_n$ behave distinctly for  $n\leq 2$ and $n\geq 3$.
Up to $O(\gamma^{-1})$, there remain only three equations as
\bb\label{trunc}
0 &=& \partial_t c_0 + \left[D + \left(\sqrt{T}\right)'\right] c_1  ,\nn \\
0 &=& \gamma c_1 + \left[\hat{D} + 2\left(\sqrt{T}\right)'\right] c_0, \\
0 &=& 3\gamma c_3 + \sqrt{6} \left(\sqrt{T}\right)' c_0~. \nn
\ee
By combining the first two equations, we get the partial differential equation
for $c_0(x,t)$ as
\bb
\partial_t c_0 &=& \left[D+\left(\sqrt{T}\right)'\right]\gamma^{-1} \left[\hat{D} + 2\left(\sqrt{T}\right)'\right] c_0 \nn\\
&=& \partial_x \left[ -\frac{f}{\gamma} +\frac{1}{\gamma}\partial_x T \right] c_0~.
\ee
By solving this equation for $c_0$, and rewriting $c_1$ and $c_3$ in terms of $c_0$ as
given in Eq.~(\ref{trunc}), we finally get the solution for $P(x,v,t)=\psi_0 \bar{P}(x,v,t)$
through Eq.~(\ref{expan}) in the large $\gamma$ limit.

In this work, we are interested in the probability distribution function of
position $x$, integrated over velocity $v$ as
\bb
\hat{P} (x,t) &=& \int_{-\infty}^{+\infty} dv~ P(x,v,t)\\ \nn
&=& \int_{-\infty}^{+\infty} dv~ \psi_0(v,T) \bar{P}(x,v,t) = c_0(x,t)
\ee
where  the orthonormal property of $\{\psi_n\}$ is used. Thus, we find
\bb\label{CorrectedOver}
\partial_t \hat{P} (x,t) =
\partial_x \left[ -\frac{f}{\gamma} +\frac{1}{\gamma}\partial_x T \right]
\hat{P} (x,t)
\ee
This result is exactly the same as the SSMD result, Eq.~(2.18) in~\cite{Sanc82}.

It is certainly different from the naive result in Eq.~(\ref{convFP}). First,
it is independent of the noise calculus, $\alpha$. Moreover, any choice of $\alpha$
in Eq.~(\ref{convFP}) is not consistent with the above equation in general. The naive result
with the anti-Ito choice ($\alpha=1$) happens to be identical to the above equation, only when $T$ is a constant~\cite{titulaer,risken81}. The Ito calculus ($\alpha=0$) also happens to give
a correct result, only when $\gamma$ is a constant~\cite{Mats00,Sekimoto}.

The correct
and general  Langevin equation corresponding to the above Fokker Plank equation,
Eq.~(\ref{CorrectedOver}), can be written as
\begin{equation}
\dot{x}=\frac{f}{\gamma }+T\left(\frac{1}{\gamma}\right)' -\alpha \left(\frac{T}{\gamma}\right)' + \sqrt{\frac{T}{\gamma }}~\xi (t)~,
\label{correc-convLang}
\end{equation}
where the noise-calculus-dependent drift force is included to cancel out
the additional drift term induced by the multiplicative noise. This inclusion
implies that the naive approach with an extra {\em physical} drift force
cannot describe the correct overdamped limit in general.

The overdamped limit  as above is an {\em extreme} limit of large $\gamma$ such that $\gamma \gg (\Delta t)^{-1}$.
However, the generalized Brinkman's hierarchy of Eq.~(\ref{Coeff}) allows us to derive a systematic expansion in terms of $\gamma^{-1}$
for $\gamma \gg 1$. Thus, the inverse friction expansion should be valid for a reasonably
large value of $\gamma$. In this case, one can not simply ignore $\partial_t c_n$ in Eq.~({\ref{Coeff}), which
makes the analysis complicated.

We again focus on deriving a partial differential equation for $c_0(x,t)$, which
is the probability distribution function $\hat{P}(x,t)$ after integrating out the velocity degree of freedom.
First, we take the hierarchy equations into the Laplace space as
\bb \label{Coeff_Lapace}
sc^L_n - c_n(0) &&= -n \gamma c^L_n -\sqrt{n+1}~ \mathcal{L}_{n+1}c^L_{n+1}    \\
&& - \sqrt{n}~\hat{\mathcal{L}}_{2n}c^L_{n-1}- \sqrt{n(n-1)(n-2)}~ \mathcal{T} c^L_{n-3} \nn ~,
\ee
where $c^L_n$ is the Laplace transform of $c_n$ as $c^L_n\equiv \int_0^\infty dt~ e^{-st} c_n(t)$
and $c_n(0)$ is the initial value at $t=0$. Here, we use simple notations as
\bb
\mathcal{L}_n&=&\sqrt{T}\partial_x +n\left(\sqrt{T}\right)' ,  \nn\\
\hat{\mathcal{L}}_n&=&\sqrt{T}\partial_x +n\left(\sqrt{T}\right)'-f/\sqrt{T} ,\\
\mathcal{T} &=& \left(\sqrt{T}\right)' . \nn
\ee

For simplicity, we assume the initial condition $c_n(0) = 0$ for $n\ge 1$, which implies
the Maxwell velocity distribution initially at local temperature $T(x)$. Then, similar to the overdamped case,
it is easy to show that $c^L_n\sim O(\gamma^{-n})$ for $n=0,1,2$ and
$c^L_n\sim O(\gamma^{-(n-2)})$ for $n\geq 3$. Collecting all terms up to $O(\gamma^{-3})$, we find
\bb\label{Laplacec0}
sc^L_0 - c_0(0)&&= \mathcal{L}_1 \frac{1}{s+\gamma} \hat{\mathcal{L}}_2 c^L_0 \nn\\
&&+ \mathcal{L}_1 \frac{1}{s+\gamma}\mathcal{L}_2 \frac{2}{s+2\gamma} \hat{\mathcal{L}}_4 \frac{1}{s+\gamma} \hat{\mathcal{L}}_2 c^L_0 \nn\\
&&+ \mathcal{L}_1 \frac{1}{s+\gamma}\mathcal{L}_2 \frac{2}{s+2\gamma} {\mathcal{L}}_3 \frac{3}{s+3\gamma} {\mathcal{T}} c^L_0
+ O(\gamma^{-4})\nn\\
&&\equiv K^L_0(s)c^L_0(s)~.
\ee
Applying the inverse Laplace transform, we can formally write the equation
for $\hat{P} (x,t)=c_0(x,t)$ as
\bb
\partial_t \hat{P} (x,t) &=& \int_0^t \D \tau~ K_0(\tau) \hat{P}(x,t-\tau)
\ee
where $K_0(t)$ is the inverse Laplace transform of the kernel $K^L_0(s)$.
Note that this differential equation is not local in time, but has the memory kernel $K_0$.
Our derivation of the first few expansion terms for $K_0$ in Eq.~(\ref{Laplacec0})
is regarded as a substantial extension of the previous result in Ref.~\cite{risken81} to a general
inhomogeneous and nonisothermal (local temperature) case. In contrast, WT~\cite{Widd89}
also studied the general case, but assumed a Smoluchowski-type differential equation which is local in time with
a time-independent evolution operator to derive higher-order corrections.

It is easy to notice that $K_0(t)$ should decay exponentially fast ($\sim e^{-\gamma t}$) even at the first order in Eq.~(\ref{Laplacec0}).
Therefore, one can utilize a Taylor expansion for $\hat{P}(x,t-\tau)$ around $\tau =0$ and by iteration we get
\bb\label{Importantc0}
\partial_t \hat{P}(x,t) &&= \left[\int_0^t \D \tau~ K_0(\tau) \right. \nn\\ \nn
& & - \left.\int_0^t \D \tau ~\tau K_0(\tau) \int_0^t \D \tau~ K_0(\tau) + \cdots\right] \hat{P}(x,t) ~\\
&& \equiv \hat{L} (x,t) \hat{P}(x,t) ~.
\ee
Note that all the memory terms are recast into a time-dependent evolution operator $\hat{L}$. Furthermore,
as $K_0^L$ is $O(\gamma^{-1})$, the above expansion can be also regarded as another inverse friction expansion
in $\gamma^{-1}$.

First, consider the lowest order in $\gamma^{-1}$. Then, $K^L_0(s)= \mathcal{L}_1 \frac{1}{s+\gamma} \hat{\mathcal{L}}_2$
and $\hat{L}=\int_0^t \D \tau~ K_0(\tau)=\int_0^t \D \tau~\mathcal{L}_1 \e^{-\gamma \tau} \hat{\mathcal{L}}_2$, which yields
\bb
\partial_t \hat{P}(x,t) = \partial_x \left[\frac{1-\e^{-\gamma t}}{\gamma} \left(-f+\partial_x T \right) \right]\hat{P}(x,t)~.
\ee
This equation reduces to the overdamped limit of Eq.~(\ref{CorrectedOver})  in the limit
of $\gamma t \gg 1$. This is a much weaker condition for large $\gamma$, compared to the extreme limit of $\gamma \Delta t \gg 1$.
However, for small $t<\gamma^{-1}$, the exponential factor produces a non-negligible correction.

For a moment, we assume that $\gamma t \gg 1$ for simplicity. As $K_0(t)$ is a function of $\gamma t$, one can safely
replace the upper integral limit by an infinity in Eq.~(\ref{Importantc0}).
Then, we get a time-independent evolution operator as
\bb\label{Generatingc0}
\hat{L}(x) &=& K^L_0(0) + \left.\partial_s K^L_0(s)\right|_{s=0} K^L_0(0)+ \cdots  ~.
\ee
Taking the terms up to the second order in the Taylor expansion, it is easy to see that the equation is valid up to $O(\gamma^{-3})$.
Finally, using Eq.~(\ref{Laplacec0}), the evolution operator  is written, up to $O(\gamma^{-3})$, as
\bb
\hat{L} (x) &=& \mathcal{L}_1 \frac{1}{\gamma} \hat{\mathcal{L}}_2
+ \mathcal{L}_1 \frac{1}{\gamma}\mathcal{L}_2 \frac{1}{\gamma} \hat{\mathcal{L}}_4 \frac{1}{\gamma} \hat{\mathcal{L}}_2 \nn\\
&+& \mathcal{L}_1 \frac{1}{\gamma}\mathcal{L}_2 \frac{1}{\gamma} {\mathcal{L}}_3 \frac{1}{\gamma} {\mathcal{T}}
- \mathcal{L}_1 \frac{1}{\gamma^2}\hat{\mathcal{L}}_2\mathcal{L}_1 \frac{1}{\gamma} \hat{\mathcal{L}}_2 ~.
\ee
Rearranging the above terms reproduce the WT result in Eq.~(3.1) of Ref.~\cite{Widd89}. It is quite straightforward to
obtain higher-order terms in $\gamma^{-1}$ in a Smoluchowski-type expression for $\gamma t\gg 1$ and also feasible to
obtain a time-dependent evolution operator $\hat{L} (x,t)$ in higher orders for $\gamma \gg 1$.

We now want to confirm and test the robustness of Eq.~(\ref{CorrectedOver}) by numerical simulations
for simple examples.
First, we perform numerical integrations of the second order SDE, Eq.~(\ref{2ndSDE}),
for large $\gamma$. Casting the second order SDE into a set of two first order SDE's
and integrating them during time interval $[t, t+\Delta t]$, we get
\bb
\Delta x&=& v \Delta t~, \nonumber\\
\Delta v &=& \left(-\gamma v + f\right)\Delta t +  (\sqrt{T\gamma}~)_{\textrm I}~\Delta W~,
\label{SDE_Kramers}
\ee
where we set $m=1$ and choose the Ito calculus for convenience without loss of generality
for small $\Delta t$. Here, we take $\Delta t=10^{-3}$ and the initial distributions are Gaussian with variance $2$ centered on $x=1$ for the position and centered on $v=0$ for the velocity. To obtain a reasonable accuracy for the probability distribution function, we repeat simulations for $2\sim5\times 10^6$ samples.

Next we perform numerical simulations, using our result of Eq.~(\ref{correc-convLang}) with
$\alpha=0$ (identical with any other choice of $\alpha$) as
\begin{equation}
\Delta x=\left[\frac{f}{\gamma}+T\left(\frac{1}{\gamma}\right)'\right]\Delta t+\left(\sqrt{\frac{T}{\gamma}~}\right)_{\textrm I}~\Delta W~,
\label{SDE_1}
\end{equation}
and also using the naive result of Eq.~(\ref{convSDE})
\begin{equation}
\Delta x=\left[\frac{f}{\gamma}+\alpha \left(\frac{T}{\gamma}\right)'\right]\Delta t+\left(\sqrt{\frac{T}{\gamma}~}\right)_{\textrm I}~\Delta W~.
\label{SDE_2}
\end{equation}
Finally we compare the results from Eq.~(\ref{SDE_Kramers}) with those from Eq.~(\ref{SDE_1}) and Eq.~(\ref{SDE_2}).

In the first example, we take $\gamma(x) = \gamma_0(1+ \e^{-x^2/2})$
and $T(x) =2/[(1+ \e^{-x^2/2})(1+2x^2)^2]$ with $f(x)=0$.
Here, we set $\gamma_0=10$ which is much smaller than $(\Delta t)^{-1}=10^{3}$, but
still reasonably good for the first-order approximation in $\gamma^{-1}$. Thus, this example should be
well described by Eq.~(\ref{CorrectedOver}) at $t=5$ $(>\gamma_0^{-1}=0.1)$.
In Fig.~\ref{abb2}, one can
easily see that the naive overdamped limit with either $\alpha=1$ or $\alpha=0$ does not fit the data points obtained from Eq.~(\ref{SDE_Kramers}), though the latter seems to fit better by chance. In contrast, our overdamped limit given by
Eq.~(\ref{SDE_1}) shows an excellent agreement.
%%++++++++++++++++++++++++++++++++++++++++++++++++++++++++++++++++++++++++++++++++++++
\begin{figure}
\centering
\includegraphics*[width=\columnwidth]{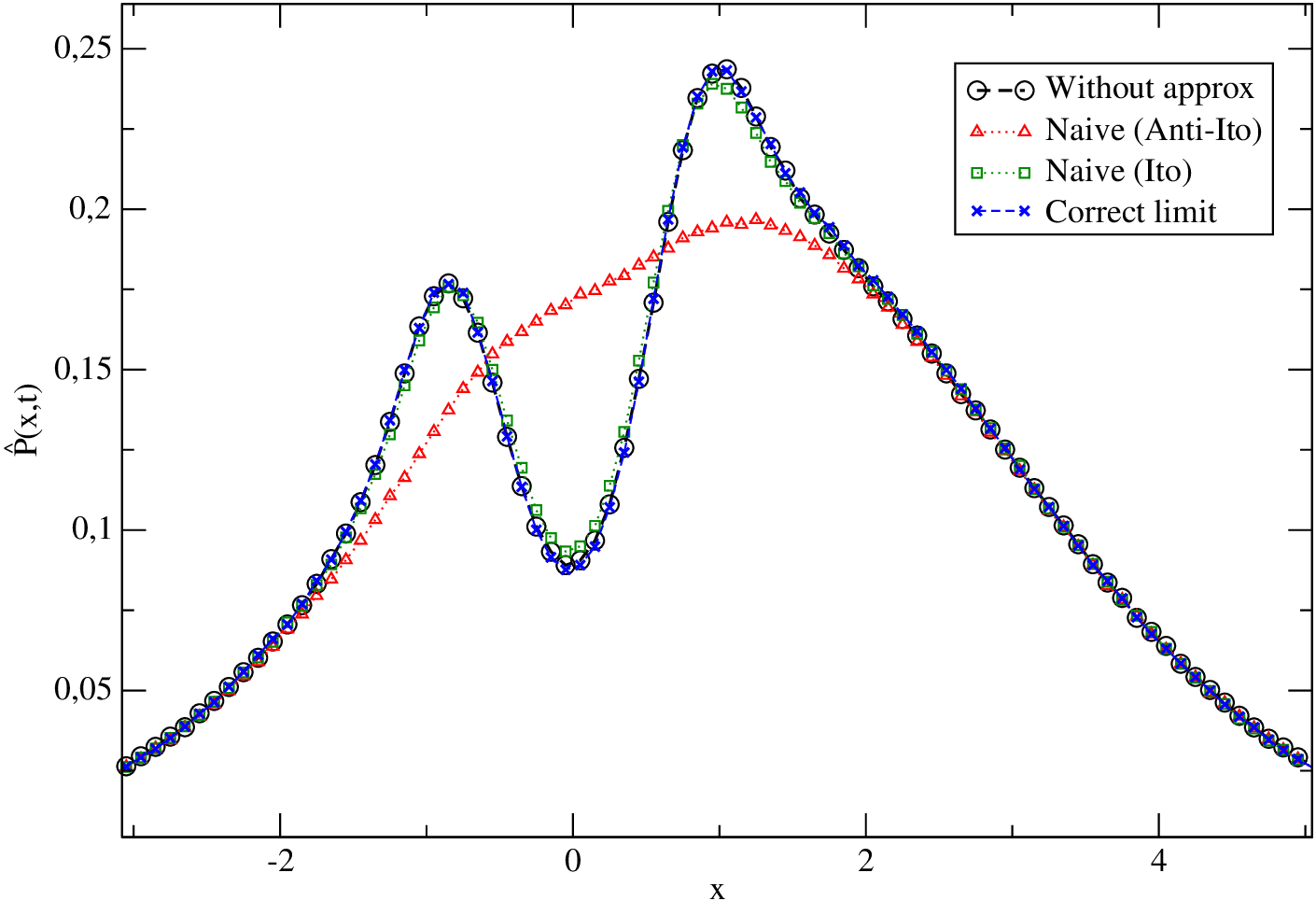}
\caption{Probability distribution function $\hat{P}(x,t)$ at $t=5$. We take
$\gamma(x) = \gamma_0(1+ \e^{-x^2/2})$, $T(x) =2/[(1+ \e^{-x^2/2})(1+2x^2)^2]$, and $f(x)=0$
with a large value of $\gamma_0=10$.
 Circles (without approx.) and crosses (correct limit) represent the data obtained from Eq.~(\ref{SDE_Kramers}) and Eq.~(\ref{SDE_1}), respectively, which overlap each other
  very well. Squares (Ito) and triangles (anti-Ito) represent the data obtained from Eq.~(\ref{SDE_2}) with $\alpha=0,~1$, respectively.}
\label{abb2}
\end{figure}
%%++++++++++++++++++++++++++++++++++++++++++++++++++++++++++++++++++++++++++++++++++++

In the second example, we take $\gamma(x) = \gamma_0(1 + \frac{4.4x}{3(x^2+1)})$
and $T(x) =(3+\frac{4x}{x^2+1})^2/[4(3+ \frac{4.4x}{x^2+1})]$ with $f(x)=-2x$.
We take $\gamma_0=30$ which should be large enough for Eq.~(\ref{CorrectedOver}) at $t=10$.
Again, the data in Fig.~\ref{abb3} show a perfect agreement between
our overdamped limit given by Eq.~(\ref{SDE_1}) and the stochastic differential
equation of Eq.~(\ref{SDE_Kramers}).
%%++++++++++++++++++++++++++++++++++++++++++++++++++++++++++++++++++++++++++++++++++++
\begin{figure}
\centering
\includegraphics*[width=\columnwidth]{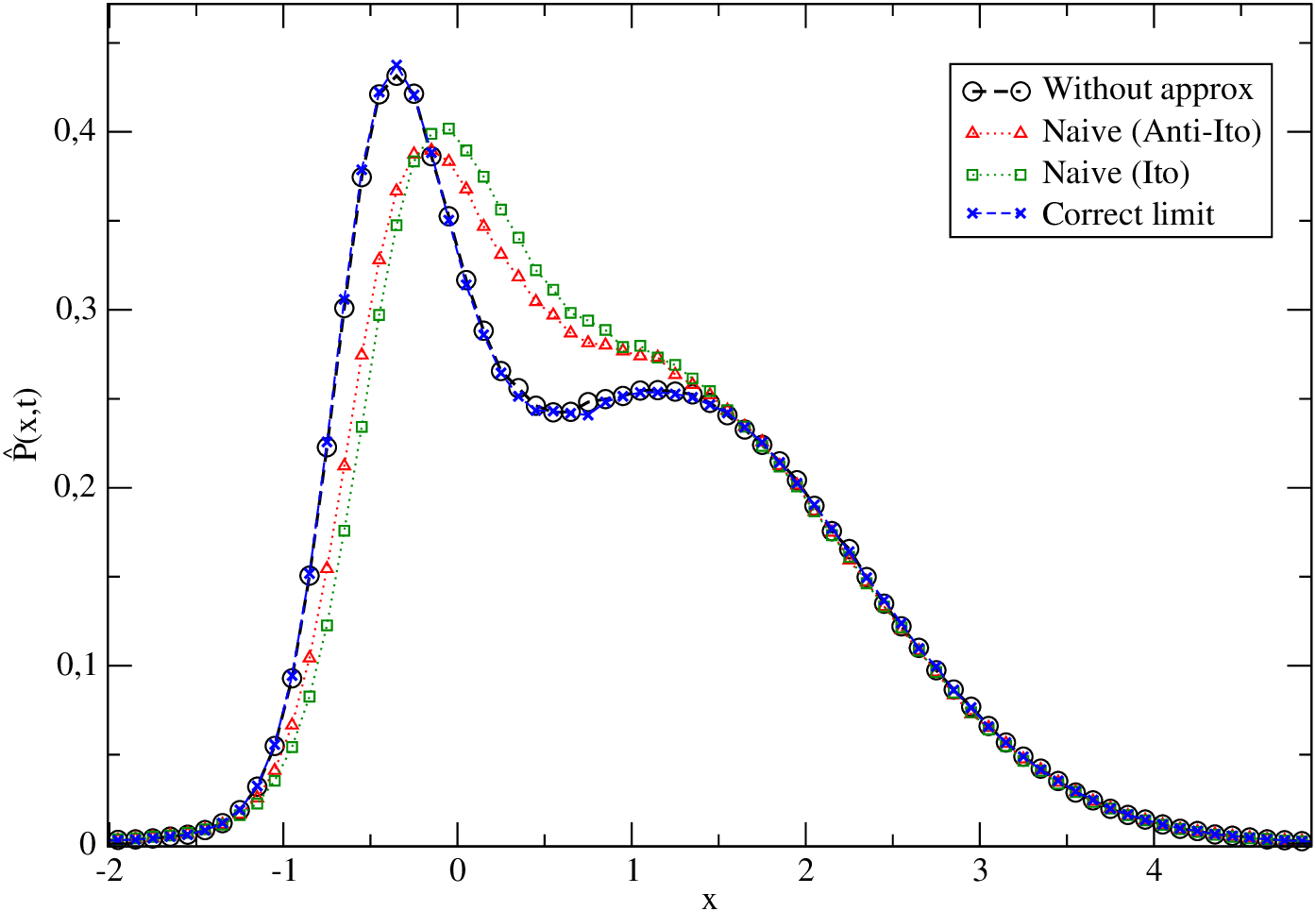}
\caption{Probability distribution function $\hat{P}(x,t)$ at $t=10$. We take
$\gamma(x) = \gamma_0(1 + \frac{4.4x}{3(x^2+1)})$,
$T(x) =(3+\frac{4x}{x^2+1})^2/[4(3+ \frac{4.4x}{x^2+1})]$, and $f(x)=-2x$
with $\gamma_0=30$. The same symbols are used as in Fig.~\ref{abb2}.
\label{abb3}}
\end{figure}
%%++++++++++++++++++++++++++++++++++++++++++++++++++++++++++++++++++++++++++++++++++++

Finally, we discuss the mass zero limit of Eq.~(\ref{2ndSDE}). Ao {\it et al.} derived the Fokker Planck equation in the mass zero limit when $T$ is a constant~\cite{Ao07}. Here, we extend their result to the general case where $T=T(x)$. Starting from
the Kramers equation given by Eqs.~(\ref{Kramer.rev.irr}) and (\ref{rev.irr}), we change the variables such that $s=t/\sqrt{m}$ and $u=v\sqrt{m}$ to obtain the covariant form
of the Kramers equation in terms of variables $(x,u,s)$ as
\bb
\partial_s P(x,u,s) = \left(L_{rev}+L_{irr}\right) P(x,u,s)
\ee
with
\bb
L_{rev} = -u \partial_x - f \partial_u, \quad
L_{irr} = \gamma_m \partial_u \left[ u+ T \partial_u\right]~,
\ee
with $\gamma_m=\gamma/\sqrt{m}$.
These equations are the same as Eqs.~(\ref{Kramer.rev.irr}) and (\ref{rev.irr})
by replacing $\gamma$ by $\gamma_m$ and setting $m=1$. It is obvious that the mass zero limit ($m\rightarrow0$) is equivalent to the large $\gamma_m$ limit as long as
$\gamma$ does not vanish. Thus, one can perform exactly the same transformation as we did with the variable $\gamma_m$ and take the large $\gamma_m$ limit to obtain
\bb\label{CorrectedMass}
\partial_s \hat{P} (x,s) =
\partial_x \left[ -\frac{f}{\gamma_m} +\frac{1}{\gamma_m}\partial_x T \right]
\hat{P} (x,s) ~.
 \ee
By returning back to the original variables of $(x,t)$, we can easily recover Eq.~(\ref{CorrectedOver}). This proves that the mass zero limit is equivalent to
the overdamped limit for general cases.
We check this result by numerical simulations when $T$ is not a constant
with $\gamma(x) = 1 + \e^{-x^2/2}$, $T(x)=1/[2(1 + \e^{-x^2/2})(x^2+1)^2]$, and
$f(x)=-x/5$ for small $m=0.01$.
%%++++++++++++++++++++++++++++++++++++++++++++++++++++++++++++++++++++++++++++++++++++
\begin{figure}
\centering
\includegraphics*[width=\columnwidth]{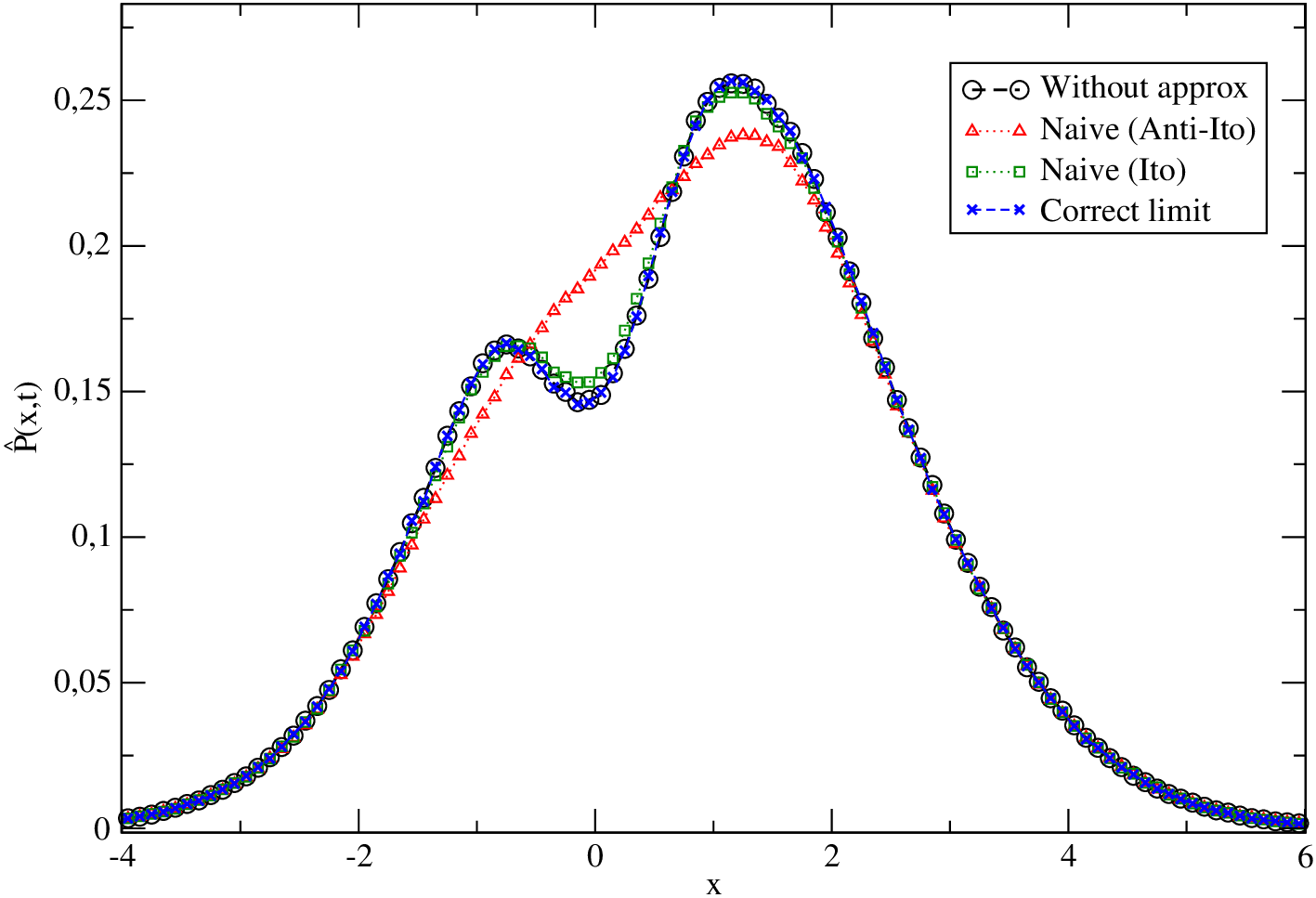}
\caption{Probability distribution function $\hat{P}(x,t)$ at $t=1$. We take
$\gamma(x) = 1 + \e^{-x^2/2}$, $T(x)=1/[2(1 + \e^{-x^2/2})(x^2+1)^2]$, and
$f(x)=-x/5$ with a small value of $m=0.01$.
The same symbols are used as in Fig.~\ref{abb2}.
\label{abb4}}
\end{figure}
%%++++++++++++++++++++++++++++++++++++++++++++++++++++++++++++++++++++++++++++++++++++
Again we have an excellent agreement between the simulations on the original
second order SDE and our equation in the mass zero limit.

To summarize, we derive the overdamped Fokker Planck equation
for the Brownian motion in a general inhomogeneous medium with a position-dependent
friction coefficient as well as a position-dependent temperature. Our result is
consistent with the SSMD result~\cite{Sanc82} and at the next order with the WT result~\cite{Widd89}.
Our derivation procedure is
straightforward and allows for a systematic calculation of higher-order
corrections  without any assumption. We also show that the mass zero limit is generally equivalent to the overdamped limit in
one space dimension.
We may note that this procedure is a direct derivation from the underdamped Kramers equation, however on other systems, the overdamped equation may not be simply the limit of an underdamped one \cite{Sanc00}.

We thank Ping Ao for interesting discussions, and Shin-ichi Sasa and Fabio Marchesoni for helpful references. We also thank Kunimasa Miyazaki  for stimulating communications.  This research was supported by the NRF Grant No.~2013R1A1A2011079(C.K.) and 2013R1A1A2A10009722(H.P.).


\begin{thebibliography}{99}
\bibitem{exp} The mass value $m$ can be easily restored back into any equation or result by simply replacing
$f$ by $f/m$, $\gamma$ by $\gamma/m$, and $T$ by $T/m$.
\end{thebibliography}
\end{document}